\documentclass[aip,jcp,reprint]{revtex4-1}
\usepackage{hyperref}
\usepackage{subeqnarray}
\usepackage{comment}
\usepackage{amsmath,amssymb,bbm}
\usepackage{graphics,graphicx}
\usepackage{mathrsfs}
\usepackage{dsfont}
\usepackage{cleveref}
\usepackage[dvipsnames]{xcolor}
\usepackage{epstopdf}
\usepackage{multirow}
\usepackage{booktabs}
\usepackage[english]{babel}

\newcommand{\Dop}{\hat{D}}
\newcommand{\Dm}{D}
\newcommand{\trace} [1]{\trm{Tr}\{#1\}}
\newcommand{\nocc}{\rho}
\newcommand{\Fop}{\hat{H}}
\newcommand{\Pop}{\hat{P}}
\renewcommand{\Im}{I}

\newcommand{\FD} {\rho_{\beta}}
\newcommand{\deltaFD} {\delta_{\beta}}
\newcommand{\deltaP} {\omega_{\beta}}
\newcommand{\deltaH} {\bar{\omega}_{\beta}}
\newcommand{\deltaPn} {\omega}
\newcommand{\deltaHn} {\bar{\omega}}
\newcommand{\DeltaP} {\hat{\Omega}_{\beta}}
\newcommand{\DeltaH} {\hat{\bar{\Omega}}_{\beta}}
\newcommand{\DeltaPn} {\hat{\Omega}}
\newcommand{\DeltaHn} {\hat{\bar{\Omega}}}

\newcommand{\epshomo} {\epsilon_{N}}
\newcommand{\epslumo} {\epsilon_{N+1}}
\newcommand{\epsgap} {\Delta}

\newcommand{\epsPP} {\langle\epsilon_{\beta}\rangle}
\newcommand{\epsHH} {\langle\bar{\epsilon}_{\beta}\rangle}

\newcommand{\dosfixed} {g}
\newcommand{\dosgap} {f}
\newcommand{\kB}{k_\trm{B}}

\newcommand*{\trm}{\textrm}
\newcommand*{\eq}  [1]{Eq.~(\ref{#1})}
\newcommand*{\eqs}[1]{Eqs.~(\ref{#1})}

\newcommand*{\fig}  [1]{Fig.~\ref{#1}}
\newcommand*{\figs} [1]{Figs.~\ref{#1}}
\newcommand*{\reff}[1]{(\ref{#1})}

\newcommand*{\colb}[1]{{\color{black}{#1}}}

\newcommand*{\colg}[1]{{\color{black}{#1}}}
\newcommand*{\colp}[1]{{\color{black}{#1}}}

\begin{document}

\title{Gap  edge eigenpairs from density matrix purification using moments of the Dirac distribution}

\author{Lionel A. Truflandier}
\email{lionel.truflandier@u-bordeaux.fr}
\affiliation{Université de Bordeaux, CNRS, Bordeaux INP, Institut des Sciences Moléculaires, UMR 5255, F-33400 Talence, France}

\date{\today}
\begin{abstract}
In this work, we propose a simple method to resolve the eigenstates located at the band gap edges of an electronic eigenspectrum 
using only the quasi-purified one-particle density matrix as input. The theoretical framework relies on the decomposition of the 
occupation number variance into a particle and a hole moment. These moments, when purified using power narrowing iterations, \colp{allow us} 
to isolate the \colp{highest} occupied and \colp{lowest} unoccupied single-state projectors, \colp{giving ready access} to the corresponding eigenpairs. 
We demonstrate that when degeneracy is encountered, power narrowing remains able to deliver relevant mixed states. From a benchmark 
of selected molecules, we show that the method is robust and efficient since it requires \colp{from four to sixteen matrix-matrix multiplications
in  the worst case}.  The possibility of reducing the computational cost using the Lanczos subspace approach is discussed. The \colp{algorithm
simplicity} of power narrowing makes it very easy to implement in electronic structure codes or libraries already featuring Fermi operator 
expansion or density matrix purification.
\end{abstract}
\maketitle
Density matrix purification\cite{niklasson_expansion_2002,niklasson_trace_2003,rubensson_density_2008,truflandier_communication_2016} 
(DMP) and Fermi operator ---Chebyshev polynomials--- expansion\cite{goedecker_tight-binding_1995,liang_improved_2003,liang_fast_2004} 
(FOE) combined with sparse matrix-multiply (MM) are now well-established routines in the 
arsenal to achieve linear scaling electronic structure calculations\cite{bowler_methods_2012},
their incorporation into standard codes being facilitated by the development of libraries 
specially dedicated to sparse density matrix solvers\cite{mohr_efficient_2017,dawson_massively_2018,fattebert_hybrid_2024}.
A renewed interest \colp{in} DMP has emerged towards the conventional cubic scaling methods 
thanks to the \colp{graphics} processing unit (GPU) by taking full advantage of the accelerated dense MMs\cite{cawkwell_computing_2012,cawkwell_computation_2014,finkelstein_fast_2023}.
Recent works shifting towards advanced artificial intelligence hardware, and
resting on mixed precision arithmetic\cite{dawson_reducing_2024}, have reported impressive speed-up 
with respect to standard GPUs\cite{finkelstein_mixed_2021,pederson_large_2023}. However, the downside compared 
to diagonalization is that DMP does not offer a systematic route to the eigenstates.
Band gap edge eigenstates, \colp{of} paramount importance for the so-called higher-occupied (HO) 
and lower-unoccupied (LU) levels, are fundamental descriptors of the electronic structure of a system. 
For ground state properties, they are usually identified by applying the Aufbau principle to the Hamiltonian matrix 
eigenspectrum. This poses challenges to standard symmetric matrix diagonalisation algorithms when 
the size of the problem reaches several tens \colp{of} thousands, even for the Krylov subspace methods\cite{liesen_krylov_2012,sleijpen_jacobi-davidson_2000,rohrig-zollner_increasing_2015,lee_robust_2017}
based on earlier Lanczos\cite{lanczos_iteration_1950,paige_computational_1972} or Davidson\cite{davidson_iterative_1975,morgan_generalizations_1986} 
works, where only a few of the extreme eigenvalues are accessible at an affordable cost. To resolve interior eigenpairs inscribed 
in a domain several methods were developed.
They are essentially built on the principle to filter the eigenspectrum around a target value\cite{ericsson_spectral_1980,morgan_computing_1991,wang_electronic_1994,xiang_linear_2007} 
---the Fermi level for instance--- 	or an energy domain\cite{fang_filtered_2012}, and solve for eigenpairs using the aforementioned Krylov methods\cite{ericsson_spectral_1980,bekas_computation_2008,zhou_self-consistent-field_2006,fang_filtered_2012,lee_robust_2017}.  
Other approaches rely on the Green function operator expansions\cite{chen_general_1996,mandelshtam_quantum_1997,mandelshtam_low-storage_1997,yu_calculation_1997,yu_spectral_1999} or
Green function projected contour integration\cite{sakurai_projection_2003,ikegami_filter_2010}.

It turns out that \colp{the} DMP process can also be used to extract interior eigenstates. Rubensson et al. showed that 
the intermediate density matrices can be recycled to locate the HO/LU (density matrix) eigenvalues\cite{rubensson_computation_2008,rubensson_interior_2014}
at no cost. Extraction of the eigenvectors is performed in a second step by combining the shift-and-square method\cite{wang_electronic_1994} 
with Lanczos iterations\cite{kruchinina_--fly_2018}. As a result, the computational overhead is
governed by the matrix-vector multiplications. In this work, inspired by the work of Rubensson and the seminal paper by Daw\cite{daw_model_1993},
we propose a different approach where we build filters from an nearly converged density matrix to extract subspaces around the
HO and LU levels and either purify them \colp{to access} eigenprojectors or apply the Lanczos
method  \colp{to access} eigenvectors. \colp{From these, we can directly compute the corresponding eigenvalues.}
In the first case, this implies dense \colp{matrix-matrix} mutiplications but the method is fairly simple to implement and independent of the purification polynomials.

\begin{figure*}[!hptb]
  \centering
  \includegraphics[scale=0.75]{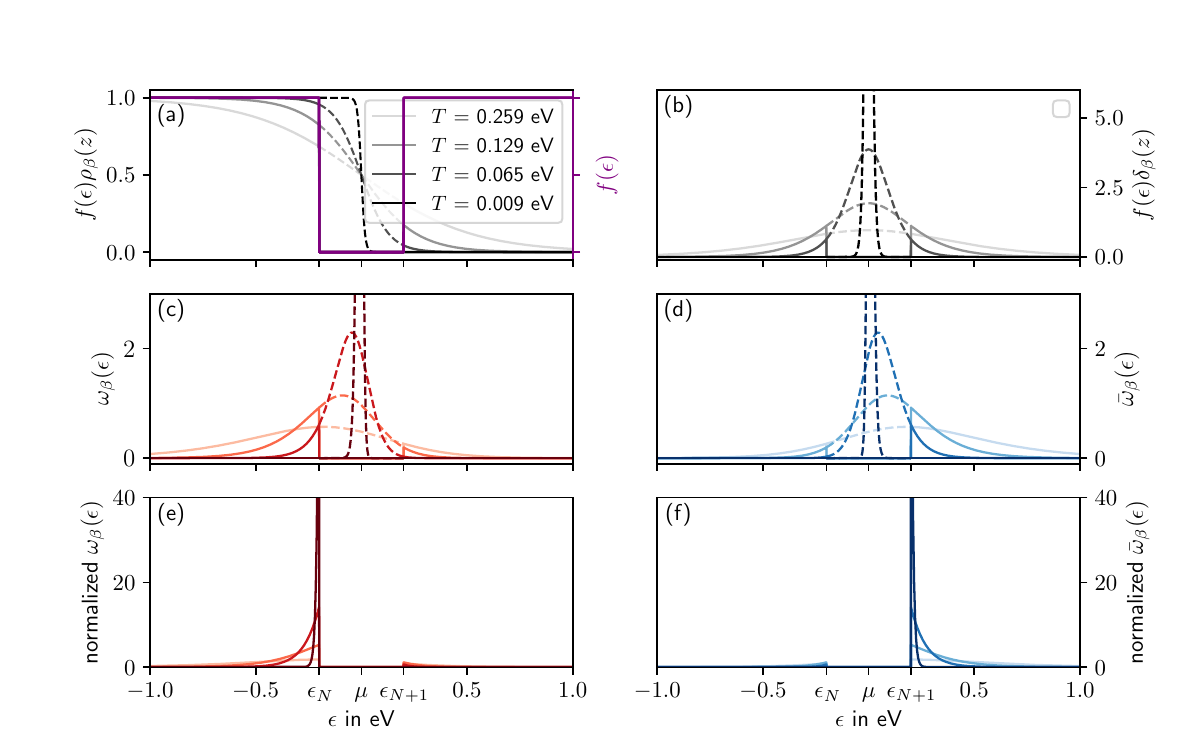}  
  \caption{
   (a) \colg{Product of the Fermi-Dirac $\FD$-distribution of \eq{eq:fd_function}  with the double step function $f(\epsilon)$ of  \eq{eq:g101}} as a function of the energy,
    for different temperatures in units of $\kB$. $f(\epsilon)$ is plotted with a solid purple line. The band gap is fixed to $0.4$ eV 
    and centered on the chemical potential $\mu=0$. (b) Corresponding Dirac $\deltaFD$-distributions. (c) and (d) Particle and hole filters, 
    $\deltaP$ (red colormap) and $\deltaH$ (blue color map), respectively.  \colg{In these 4 plots the dashed lines refer to the expected 
    behavior without band gap, that is, when the density of states is fixed: $g(\epsilon)=1$}. (e) and (f) Normalized $\deltaP$ and $\deltaH$ as a function of temperature. 
    Positions of the last occupied ($\epshomo$), first unoccupied ($\epslumo$) and $\mu$ energies are indicated on the $x$-axis.} 
  \label{figure1}  
\end{figure*}

Let's begin with the effective Hamiltonian operator $\Fop$ of size \colg{$M\times M$}, and its set of \colg{$M$} orthonormalized 
one-electron eigenvectors $\{|\psi_i\rangle\}$ and eigenvalues $\{\epsilon_i\} $, and consider the canonical inverse temperature  
\colg{$\beta=1/T$, where $T$ is in units of the Boltzmann's constant.} \colg{For the restricted case with a} \colg{$2N$}-electron system, \colg{where $M>N$}, 
the expression of the spinless one-particle density operator is given by
\colg{\begin{equation}
   \Dop_\beta \equiv \sum^M_{i=1} \nocc_\beta(\epsilon_i)\Pop_i  \label{eq:densop}
\end{equation}}
where $\Pop_i$ is a single-state projector
\begin{equation}
   \Pop_i \equiv  |\psi_i\rangle\langle\psi_i| 
   \label{eq:projop}
\end{equation}
%
and $\nocc_\beta(\epsilon_i)$ is the occupation number of the state at energy $\epsilon_i$. In the 
basis of the eigenstates, the density matrix is diagonal with matrix elements 
\colg{$\langle\psi_i|\Dop_\beta|\psi_i\rangle=\nocc_\beta(\epsilon_i)$}. At thermodynamic equilibrium, 
the occupation numbers are obtained from the Fermi-Dirac distribution
\begin{equation}
\label{eq:fd_function} 
	\rho_\beta(\epsilon_i)\equiv\left( 1+e^{-\beta(\mu-\epsilon_i)} \right)^{-1} 	
\end{equation}
for which \colg{$\nocc_\beta(\epsilon_i) \in[0,1]$} is verified. 
The chemical potential $\mu$ is chosen
to ensure \colg{$\trace{\Dop_\beta}=\sum^M_i\nocc_\beta(\epsilon_i) =N$}. \colg{For the unrestricted case, provided 
two components for $\mu$, the same relations hold independently for the two spin channels}. Combining \eqs{eq:densop} 
and \reff{eq:fd_function} \colg{along with the spectral decomposition of $\Fop$}, we obtain a one-to-one correspondence
between $\Dop$ and $\Fop$, without requiring the knowledge of the eigenvectors, 
\begin{equation}
	 \Dop_\beta \equiv \left( \Im + e^{-\beta(\mu\Im - \Fop)} \right)^{-1} 
	 \label{eq:fermidirac} 
\end{equation}
This is the starting point for the development of the DMP and FOE methods, 
trying to resolve $\Dop$ from $\Fop$ by a polynomial recursion or expansion. 
For the sake of demonstration, we shall consider a very dense spectrum of
\colg{eigenvalues} in order to approach a continuous distribution of occupation numbers,
characterized by a constant density of states $\dosfixed(\epsilon)=1$.
Complying with the notation of Daw\cite{daw_model_1993}, 
the continuous version of \eq{eq:fd_function} reads
\begin{equation}
\label{eq:fd} 
	\FD(z)\equiv\left( 1+e^{-\beta z} \right)^{-1} 
\end{equation}
with $z=\mu-\epsilon$. By taking the derivative of $\FD$ with respect to $z$, we obtain something 
known as the occupation variance,
\begin{equation}
\label{eq:fd_dirac} 
	 \deltaFD(z)\equiv \beta\FD\left(1-\FD\right)
\end{equation}
which behaves very much like a Dirac distribution. Actually, $\deltaFD$
has all the required properties: positive definiteness, even symmetry, and
\begin{equation}
\label{eq:lim_fd_dirac}
\int^{+\infty}_{-\infty}\deltaFD(z) d\epsilon=1, \ \ \trm{with}\ \ \    	
{\operatorname{\lim_{\beta\rightarrow\infty}\deltaFD}}\: 
	\begin{cases}
		+\infty\;\text{if}\: \epsilon = \mu \\ 
         0\ \ \ \ \ \text{if} \: \epsilon \neq \mu 
   	\end{cases}
\end{equation}
Note that from here  to the end analytical integrals 	are computed in the infinite interval $(-\infty,+\infty)$. 
The spread of this distribution, measured through the standard deviation, is equal to $\pi/(\beta\sqrt{3})$. 
Note that the chemical potential is obtained by
\begin{equation}
\label{eq:mu_fd_dirac}
	\mu=\int \epsilon\;\dosfixed(\epsilon)\deltaFD(z) d\epsilon
\end{equation}
On \figs{figure1}(a) and (b), in dashed lines, $\FD$ and $\deltaFD$ are plotted for different temperatures.
We already see that the latter can be used to filter the eigenspectrum around $\mu$.
The central point of this paper is that $\deltaFD$ can be decomposed into a sum 
of two moments of order 2, using
\begin{equation}
\label{eq:mom_fd_dirac}
	 \deltaFD = \beta\FD^2(1-\FD) +  \beta\FD(1-\FD)^2	 
\end{equation}
\colg{As illustrated by the dashed lines in \figs{figure1}(c) and (d),  we observe that the first and second terms on the right-hand side (rhs) 
of \eq{eq:mom_fd_dirac} also behave like Dirac $\delta_\beta$-distributions. Although they are not symmetric, the two  
distributions are related by mirror symmetry.} From \eq{eq:fd}, we can show \colg{analytically} that
\begin{equation}
\label{eq:mom_int}
	\beta\int\dosfixed(\epsilon)\FD^2(1-\FD)d\epsilon =  \beta\int\dosfixed(\epsilon)\FD(1-\FD)^2 d\epsilon = \frac{1}{2}
\end{equation}
and more importantly,
\begin{subequations}
\label{eq:mom_int_e}
\begin{align}
	\beta\int \epsilon\;\dosfixed(\epsilon)\FD^2(1-\FD) d\epsilon&=\frac{\mu}{2}  - \frac{1}{2\beta} \label{eq:mom_int_e_a}\\
	\beta\int \epsilon\;\dosfixed(\epsilon)\FD(1-\FD)^2  d\epsilon&=\frac{\mu}{2} + \frac{1}{2\beta} \label{eq:mom_int_e_b} 		
\end{align}
\end{subequations}
If we then renormalize \eqs{eq:mom_int_e_a} and \reff{eq:mom_int_e_b} using \eq{eq:mom_int}, 
we have an expression for the mean energies:
\begin{subequations}
\label{eq:momprop2}
\begin{align}
	\epsPP&\equiv \mu - \frac{1}{\beta} \label{eq:mom_intn_e_a}\\
	\epsHH&\equiv \mu +\frac{1}{\beta} \label{eq:mom_intn_e_b} 	
\end{align}
\end{subequations}
%
%
%
%
%
%
Now, let's consider a non-uniform distribution by introducing a band gap as defined in the 
following double step function
\begin{equation}
\label{eq:g101}
{\dosgap(\epsilon)}\: 
	\begin{cases}
	 1\ \ \text{if}\ \ \epsilon\in[-\infty;\epshomo]  \\ 
	 0\ \ \text{if}\ \ \epsilon\in]\epshomo; \epslumo[  \\ 
     1\ \ \text{if}\ \ \epsilon\in[\epslumo; +\infty]
  	\end{cases}
\end{equation}
where $\epshomo$ and  $\epslumo$ are the highest occupied (HO) and lowest unoccupied (LU) state energies, respectively,
and the band gap is $\epsgap= \epslumo - \epshomo$.  \colg{The product of this rectangular function with the Dirac distribution 
of \eq{eq:fd_dirac} is shown in solid lines on \fig{figure1}(b) for different temperatures. For reference, dashed lines 
correspond to constant density of states ie., without band gap.  From \eq{eq:mom_fd_dirac}, we can separate the contributions. 
They are shown in \figs{figure1}(c) and (d) using the following definitions:}
\begin{subequations}
\begin{align}
	\deltaP(\epsilon)&\equiv\beta\dosgap(\epsilon)\FD^2(1-\FD) \\
	\deltaH(\epsilon)&\equiv\beta\dosgap(\epsilon)\FD(1-\FD)^2
\end{align}
\end{subequations}
We designate them as the particle and hole moments, respectively.
\colg{From \figs{figure1}(c) and (d), we see that the maxima of $\deltaP$ and $\deltaH$ are located at $\epshomo$ 
and $\epslumo$. The number of states available below or above these values is determined by the
temperature broadening.} By applying a renormalization, we obtain something very close to 
what we need: two $\delta$-like filters located at $\epshomo$ and $\epslumo$; cf. \figs{figure1}(e) and (f). 
As a result, we can assume that good estimates of the gap edge energies are provided by
\begin{subequations}
\label{eq:momeigs}
\begin{align}
	\langle\epshomo\rangle &= \frac{\int \epsilon\;\deltaP(\epsilon) d\epsilon}{\int\deltaP(\epsilon)d\epsilon} \\	
	\langle\epslumo\rangle  &= \frac{\int \epsilon\;\deltaH(\epsilon) d\epsilon}{\int\deltaH(\epsilon)d\epsilon}	
\end{align}
\end{subequations}
Note that, at finite values of $\beta$, the chemical potential remains well-defined if \eq{eq:mu_fd_dirac} is
also properly normalized, namely
\begin{equation}
\label{eq:mu_fd_dirac_gap}
\mu = \frac{\int \epsilon\;\dosgap(\epsilon)\deltaFD(z)d\epsilon}{\int\dosgap(\epsilon)\deltaFD(z)d\epsilon}
\end{equation}
This expression is the continuous version of Daw's chemical potential.\citep{daw_model_1993}
\colp{The accuracy of these estimates depends} upon the band gap and the temperature, more precisely, they
depend on the overlap between the step function and the moments. This leads to two ill-behaved cases:
(1) at the zero-temperature limit, when particle and hole moments collapse into $\delta$-functions,
they do not overlap with $\dosgap(\epsilon)$ any more, resulting in indeterminate energy values, (2) on the opposite
side, too broad distributions can enclose states corresponding to energies far beyond 
the band gap leading to strong deviations from the expected results. 
It should be outlined that in the limit where $\dosgap(\epsilon)$ can be assimilated to $\delta$-functions 
localized at $\epshomo$ and $\epslumo$, then the estimates of \eq{eq:momeigs} become formally exact. 

Solution \colp{of} problem (1) is found by ensuring that \colp{the} evaluation of the estimates is performed in a temperature range 
compatible with the band gap. To solve problem (2), single state purification can be performed by power narrowing.
For a given $\beta$, if we introduce
\begin{subequations}
\label{eq:mompur}
\begin{align}
\deltaPn_0 = \deltaP\quad&\trm{and}\quad\deltaHn_0 = \deltaH\\
\trm{with}\quad\deltaPn_{n+1}(\epsilon)&=\frac{\deltaPn_n^k(\epsilon)}{\int \deltaPn_n^k(\epsilon) d\epsilon} \\
\trm{and}\quad\deltaHn_{n+1}(\epsilon)&=\frac{\deltaHn_n^k(\epsilon)}{\int \deltaHn_n^k(\epsilon) d\epsilon}
\end{align}
\end{subequations}
for $k>1$, we have
\begin{subequations}
\label{eq:mompurlim}
\begin{align}
	\lim_{n\rightarrow\infty}\deltaPn_{n}(\epsilon)&=\delta(\epsilon-\epshomo)\\
	\lim_{n\rightarrow\infty}\deltaHn_{n}(\epsilon)&= \delta(\epsilon-\epslumo)
\end{align}
\end{subequations}
\colb{As demonstrated in the Sec. S1 of Supporting Information (SI),
this is easily verified analytically by approximating $\omega$ with a Gaussian distribution}.
Inserted in \eqs{eq:momeigs}, we readily obtain
\begin{subequations}
\label{eq:mompureigs}
\begin{align}
\epshomo &= \int \epsilon\;\deltaPn_{\infty}(\epsilon) d\epsilon\\
\epslumo  &= \int \epsilon\;\deltaHn_{\infty}(\epsilon) d\epsilon
\end{align}
\end{subequations}
%
\begin{figure}[!hptb]
  \centering
  \includegraphics[scale=0.7]{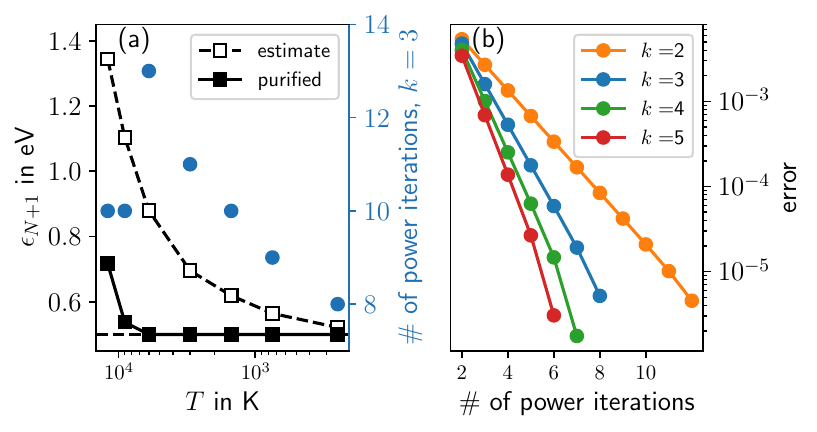} 
  \caption{(a) Estimates [\eq{eq:momeigs}] and purified [\eq{eq:mompureigs}] LU energies obtained by  power narrowing, as a function of temperature	
  \colp{on a log scale} for $\epsilon\in[-40;+40]$ and a grid spacing $\delta\epsilon=10^{-5}$ eV. Expected value is 0.5 eV. The number of iterations used in power narrowing for $k=3$ 
  is given by the blue circles (right $y$-axis). (b) Convergence of the power narrowing, for different values of $k$ at $T=250$ K,  \colp{with the error plotted on a logarithmic scale}.}
  \label{figure2}  
\end{figure}

To demonstrate the efficiency of power narrowing, numerical experiments were carried out to compute
the LU energy with a discrete representation of $\dosgap(\epsilon)$ and the moments. For this, we have generated
an evenly-spaced grid for $\epsilon$ with $\epsilon\in[-40;+40]$ with a grid spacing $\delta\epsilon=10^{-5}$ eV. 
The HO-LU gap was set to 1.0 eV centered at $\mu=0$. Recursion of \eq{eq:mompureigs} is stopped when the difference 
between two iterates falls under a threshold of $10^{-5}$ eV. Results obtained for different temperatures 
ranging from 12500 K (1.08 eV) to 250 K \colp{(0.02 eV)}  \colp{are shown} on \fig{figure2}(a) along with the number of 
iterations for $k=3$. We clearly see the benefit of power narrowing when going from estimates of
\eq{eq:momeigs} to purified energy of \eq{eq:mompureigs}. We note that at very high temperature
(above 9000 K), the purified LU energy is overestimated, but \colp{below} that it matches exactly 0.5 eV. Below
250 K, the $\deltaHn$-function collapse prevents resolution of $\epslumo$. We also observe
that higher temperature requires more iterations but the cost remains very attractive since it
increases as $O(\log(T))$. The number of iterations as a function of $k$ plotted
on \fig{figure2}(b) indicates that there is a gain in using higher powers but
with a cost in terms of floating-point multiplications, leading us to establish a trade-off.
Varying the density of states has a reasonable impact on the number of iterations, we also
found that it increases as $O(\log(1/\delta\epsilon))$. We emphasize that since $\deltaP$ 
and $\deltaH$ are localized distributions, for a fixed density of states, power narrowing
is insensitive to the spectral width of the energy.

If we go back to the discrete density operator representation of \eq{eq:densop},  and introduce
the particle and hole filters
\begin{subequations}
\label{eq:DP}
\begin{align}
\DeltaP&\equiv\Dop^2_\beta(\Im-\Dop_\beta)\\
\DeltaH&\equiv\Dop_\beta(\Im-\Dop_\beta)^2
\end{align}
\end{subequations}
we obtain the estimates of \eq{eq:momeigs} by computing
\begin{subequations}
\label{eq:mompureigsm}
\begin{align}
\langle\epshomo\rangle&=\frac{\trace{\Fop\DeltaP}}{\trace{\DeltaP}} \\
\langle\epslumo\rangle&=\frac{\trace{\Fop\DeltaH}}{\trace{\DeltaH}} 
\end{align}
\end{subequations}
By using the operator form of the recurences of \eqs{eq:mompur}, it turns out that \eqs{eq:mompurlim} 
deliver the single projectors of the HO and LU states, using
\begin{subequations}
\label{eq:DPpur}
\begin{align}
\DeltaPn_0 = \DeltaP\quad&\trm{and}\quad\DeltaHn_0 = \DeltaH\\
\DeltaPn_{n+1} = \frac{\DeltaPn_n^k}{\trace{\DeltaPn_n^k}}\quad&\trm{with}\quad\lim_{n\rightarrow\infty} \DeltaPn_{n} = \Pop_N\\
\DeltaHn_{n+1} = \frac{\DeltaHn_n^k}{\trace{\DeltaHn_n^k}}\quad&\trm{with}\quad\lim_{n\rightarrow\infty} \DeltaHn_{n} = \Pop_{N+1}
\end{align}
\end{subequations}
with $\trace{\Pop_N}=\trace{\Pop_{N+1}}=1$. \colg{We can therefore extract the eigenvalues from
\begin{subequations}
\label{eq:Deig}
\begin{align}
\epsilon_N        & = \trace{\hat{H}\Pop_N}\\
\epsilon_{N+1} &=\trace{\hat{H}\Pop_{N+1}}
\end{align}
\end{subequations}
}
Eigenvectors can be deduced from \eq{eq:DPpur}, by extracting a column and renormalizing or by a simple 
matrix-vector multiplication. The sign of the eigenvector remains undetermined, but this is not an issue. Now, the case of 
degeneracy has to be discussed. For truly degenerate states, when some of the eigenvalues of $\Dop_\beta$ 
are identical at the edges of the gap, the power narrowing will converge to a mixed state made of \colp{a} normalized 
superposition of projectors with equal weights. Therefore, the corresponding eigenvector will also be a linear 
combination of degenerate states ---note that it remains a good eigenvector with the right eigenvalue. In that case, 
the projector is not idempotent, but it is interesting to remark that it verifies $\trace{\Pop_N^2}=1/d_N$ and $\trace{\Pop_{N+1}^2}=1/d_{N+1}$, 
with $d_N$ and $d_{N+1}$, the degree of degeneracy of the HO and LU states respectively.
%
\begin{figure}[!hptb]
  \centering
  \includegraphics[scale=0.34]{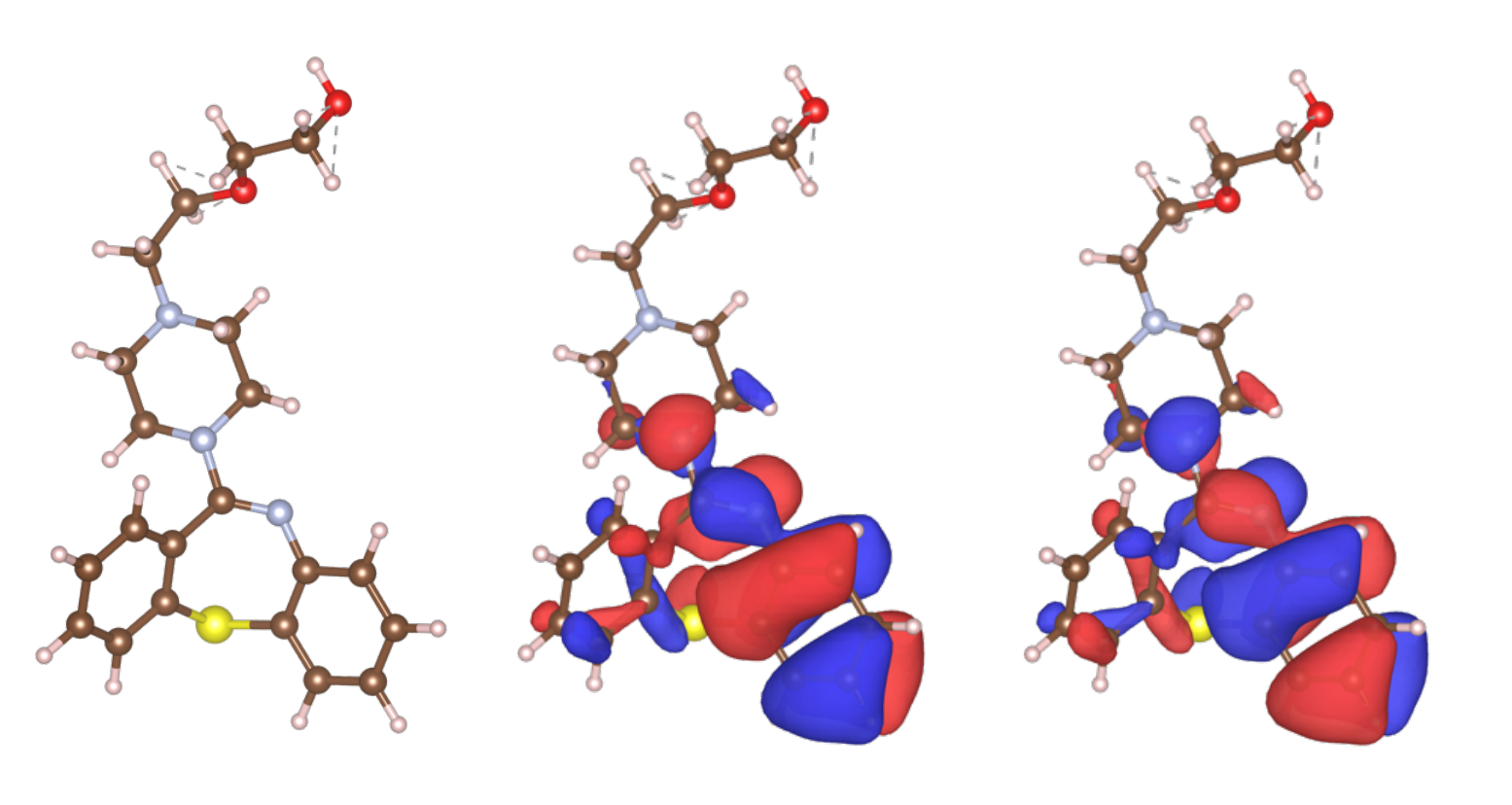}  
  \caption{From left to right: molecular structure of quetiapine, HOMO obtained with 
  power narrowing purification and diagonalisation, respectively.} 
  \label{figure3}  
\end{figure}
%
\begin{table*}[ht!]
\scriptsize
\centering
\label{tab:table1} 
\begin{tabular}{lcllrrc|rlccr|rlcrr} 
\hline\hline
  \multicolumn{7} {c}{Calculation details}    & \multicolumn{5} {c}{Higher occupied} & \multicolumn{5} {c}{Lower unoccupied}   \\
  \cline{1-7}\cline{8-12}\cline{12-17}
   Molecule & PGS &  method & basis & $M$ &  $\Delta$ &  $\tau_m\times 10^{3}$ & $d$ & state &  \multicolumn{1}{c}{err.$\langle\epshomo\rangle$} & \# P  &   \# L & $d$ &  state &
 \multicolumn{1}{c}{err.$\langle\epslumo\rangle$} &  \# P  &   \# L   \\
\hline\hline 
quetiapine$^{(a)}$&  $C_1$ &    &      &                               &           &                  &           &                  &      &   &    &          &           &                  &        &  \\
                      &                & PBE  & 6-31G$\ast\ast$  & 507   & 2.903  &0.377 & 1  & pure& -0.0033  & 4   & 5& 1 & pure & 0.0640    & 4    & 6    \\
SF$_6$        &   $O^{(b)}_h$  &          &                               &       &    &              &     &                                  &      &   &    &          &           &                  &        &  \\
                      &                & HF  & def2-SVP       &  102   &23.121  & 1.854 & 3 & mixed      & -0.9259     & 5  &   12    & 1  & pure  &     -0.2907    &  4   & 4   \\
                      &                & HF  & def2-TZVP    & 223    &22.959 & 0.836& 3 & mixed        & -0.1602     & 4   &    8     & 1  &  pure &      2.0916     &  4  & 5  \\
                      &                & HF  & def2-QZVP   & 412    &22.572 & 4.815& 3& mixed          & -0.0739    & 4   &     8 & 1    &  pure  &   16.4387      &  5  & 7    \\                     
                      &                & PBE& def2-QZVP   & 412    &7.988   &  1.024& 3 & mixed        &   0.0004    & 4   &     7   & 1    & pure  &  -0.4340    &  4   & 4  \\
                      &   $C^{(c)}_1$  &    &      &                               &           &                  &           &                  &      &   &    &          &           &                  &        &  \\
                      &                & PBE& def2-QZVP   & 412   &7.981     &1.024& 1 & pure          & 0.0002      & 16    & 11   & 1  & pure  &-0.4342   &  4    & 4  \\
C$_{60}$     &   $I^{(b)}_h$  &          &                               &&           &              &     &                      &      &   &    &          &           &                  &        &  \\
                      &                & HF    & 6-31G$\ast$   &  840   & 7.193 &  5.630 & 5 & mixed    &  -0.0460  &    4     & 5 & 3  &mixed     & 1.5409  &  4    & 7  \\
                      &                & PBE  & 6-31G$\ast$   &  840  & 1.730  &  5.576 & 5 & mixed     &  0.0033    &     4    & 5  &  3 & mixed    & -0.0011 &  4    & 4 \\
                      &  $C^{(c)}_1$ &          &                               &           &&              &     &         &           &                  &      &         &           &                  &        &  \\
                      &               & PBE  & 6-31G$\ast$    &  840  & 1.730   & 1.639 & 1 & pure  & 0.0036  & 16   & 12 &1 & pure   & -0.0010    &  14  & 8  \\
C$_{240}$  &  $C^{(c)}_1$ &    &      &                               &           &                  &           &                  &      &   &    &          &           &                  &        &  \\
                      &               & PBE  & 6-31G$\ast$    &  3360  & 1.401  & 0.751 & 1 & pure &  -0.0038 &  16   & 12 &  1 & pure  & -0.0027  & 16  & 9  \\                     
\hline\hline
\end{tabular}\\
$^{(a)}$Quetiapine geometry was optimized at the PBE/6-31G$\ast\ast$ level of theory, with the XRD structure as the initial guess. $^{(b)}$Idealized
geometries with S$-$F and C$-$C bond lengths of 1.5556 and 1.4519 \AA, respectively. $^{(c)}$Slightly distorted geometries with random variations of 
bond lengths of $10^{-4}$ \AA.
\caption{\colg{Application of power narrowing and Lanczos iterations to the HO and LU filters constructed from a density matrix sampled at $\tau_m > 10^{-4}$
. Point group symmetry (PGS) specifies whether or not symmetry constraints were applied for the calculation.
$M$ is the size of the basis set and $\epsgap$ the band gap in eV. The states of the converged $P_N$ and $P_{N+1}$ projectors are defined as 
\textit{pure} (single state) or  \textit{mixed} (superposition of degenerate states of degree $d$). Exact eigenvalues, pseudo-temperature,  
and errors in the estimates obtained using \eqs{eq:mompureigsm} —i.e., without applying power narrowing— are reported in eV.  \#P and \#L 
denote the numbers of matrix–matrix and matrix–vector multiplications required to obtain $\epshomo$ and $\epslumo$  using power narrowing 
and Lanczos, respectively. Absolute errors on the computed eigenvalues and eigenvector error norms are reported in Table S1 of the SI.}}
\end{table*}

\colb{Since power narrowing acts on the eigenvalues of the particle and hole filters, it shares the same advantages and limitations as 
density matrix purification methods when combined with the sparse matrix multiplication techniques required to achieve linear scaling.}
\colb{Whereas the method can be formally applied to isolated and extended systems,  we have reduced the scope to molecules
where the HO and LU states refers as to the frontier molecular orbitals (MO). From a large test set including various energy gaps 
and density of states, we have selected illuminating examples presented in Tab. 1. This includes the quetiapine\cite{ravikumar_quetiapine_2005}
molecule, SF$_6$ which is known to have a wide HOMO-LUMO gap, the fullerenes C$_{60}$ and C$_{240}$, both presenting a quintuply degenerate
HOMO, and triply degenerate LUMO.} Calculations were performed using restricted Hartree-Fock (HF)
and density functional theory with the Perdew-Burke-Ernzerhof\cite{perdew_generalized_1996} (PBE) exchange-correlation functional.
Pople and Weigend/Ahlrichs families of basis sets\cite{hehre_selfconsistent_1972,hariharan_influence_1973,weigend_balanced_2005} 
were used. Converged Fock matrices were generated from self-consistent-field (SCF) calculations using 
\texttt{PySCF}\cite{sun_pyscf_2018,sun_recent_2020}. Default settings for SCF convergence were retained and symmetry constraints
were imposed. Löwdin symmetric orthogonalization\cite{lowdin_nonorthogonality_1950} was applied to the Fock matrices.

In principle, any DMP or FOE density matrix ($\Dm_\beta$) can be employed to generate the hole and particle filters
of \eq{eq:DP} and perform the recursion of \eq{eq:DPpur} as long as the eigenvalue spectrum behaves like a Fermi-Dirac 
distribution and \colp{is} not too far (too close) from (to) convergence. \colb{In this work, we chose to use DMP because it does not require any information 
about the Hamiltonian interior eigenvalues and the same is true of power narrowing.}
Purification methods originally introduced for gapped systems were developed to converge rapidly towards the step function ---idempotent density matrix--- 
in such a way that temperature control was not of primary interest\cite{mniszewski_linear_2019} whereas in FOE\cite{goedecker_tight-binding_1995}, or in the recent 
wave operator minimization method\cite{leamer_positivity_2024}, the temperature can be properly defined. Fortunately, there are other quantities relevant for our study 
case, for instance the idempotency error $\tau_m=||\Dm_m - \Dm_m^2||$, with $\Dm_m$ a density matrix iterate.  \colg{Like the temperature, 
this parameter is also universal as it measures the variance of the occupation numbers at the chemical potential. Note that in the limit of low temperature,
when $\beta\Delta\gg 1$, we found\cite{truflandier_unpublished_2026} that $T/\Delta  \simeq - 1/\ln{\left(\tau_m^2/2\right)}$, directly relating
the idempotency error to the relative spectral width of the filters. A brief discussion
on this topic is given in the Sec. S3 of the SI. For all the cases investigated here, initial moments in \eqs{eq:DP} were generated form the
 \textit{last} $\Dm_m$ during the course of the DMP having $\tau_m>10^{-4}$,
ensuring that $T/\Delta> 0.05$. Power narrowing was applied using $k=3$ for the first iterate and $k=2$ for the subsequent ones 
until $||\Omega_n - \Omega_{n+1}|| < 10^{-8}$. This yields absolute errors on the eigenvalues below $10^{-10}$ eV and eigenvector error 
norms below $10^{-9}$. All errors are reported in Table S1 provided in Sec. S2 the SI.} The hole-particle canonical density matrix purification\cite{truflandier_communication_2016} (HPCP) 
was used, but we emphasize that the same trend was observed with the second-order spectral projection\cite{niklasson_expansion_2002}.

Looking first \colg{at} the HO states of the molecules, when symmetry is preserved, we found that \eq{eq:mompureigsm} already provides
good estimates of the eigenvalues and only \colp{2 or 3} iterations are necessary to reach convergence, which translates into \colp{4 or 5} MMs.
As an example, the HOMO of quetiapine extracted from $P_N$ is plotted on \fig{figure3} with the reference.
They are essentially the same with an undetermined sign for the former. The ultra wide band gap observed for SF$_6$ using HF 
and drastically reduced with PBE has no impact on the number of iterations, validating our choice for $\tau_m$. When symmetry 
is lost but quasi-degeneracy remains, the power narrowing converges to a pure state with a much larger \colp{number} of calculations, in
agreement with our previous analysis. Nevertheless, we recall that if we are searching for an eigenvalue and \colp{are} satisfied with
a mixed state then only a few iterations are necessary. The same trends can be drawn for the LU states, but not for SF$_6$ demonstrating
that \colg{estimates of \eqs{eq:mompureigsm}} are not always reliable. For the latter, explanations are found in the poor conditioning of the input matrix in \eq{eq:DPpur},
resulting from concomitant effects: the \colp{asymmetric} convergence of HPCP when dealing with low a filling factor\cite{truflandier_communication_2016}, 
and the sharp increase of the density of states around $\epslumo$ (cf.  Sec. S4 of the SI) when going from def2-SVP to def2-QZVP using HF, 
which is found to be broadened when using PBE.  As expected, given the similar \colp{density-of-states distributions} of C$_{60}$ 
and C$_{240}$, increasing the size of the problem has a weak impact on the number of iterations.

In the current state, while our method is very easy to implement and systematically convergent, it still involves \colp{repeated}
dense MMs. The standard way to reduce cost is to rely on Krylov subspace algorithms. As a preliminary step,
we applied the Lanczos method, using full reorthogonalisation, to the hole and particle filter matrices $\Omega_\beta$
and  $\bar{\Omega}_\beta$.  The size of the subspace was increased until the first eigenvalue \colp{had} converged to $10^{-8}$ eV.
Errors in the norm of the corresponding eigenvectors with respect to full diagonalisation were systematically checked and found to be below 
$10^{-10}$ \colp{(cf. Table S1 of Sec. S2)}. As presented in Tab. 1,  the number of Lanczos iterations is rather small and independent of the size or
the band gap. These results are not surprising since by construction, $\Omega_\beta$ and $\bar{\Omega}_\beta$
are already \colp{built} from a subspace of projectors. The maximum number of iterations is observed for the ill-conditioned system having 
quasi-degenerate density of states. Note that, in this case Lanczos delivers the right single state eigenvector.
We also found that other interior eigenstates are easily accessed by increasing the Krylov subspace, the limit 
being related to the eigenvalue bandwidth covered by selecting $\tau_m$. Analytical work is currently undertaken to explore 
comprehensively this option, and to introduce relevant criteria defining a secure iteration space.

To summarize, we have introduced a simple and robust method to access the gap edge eigenvalues and eigenvectors 
at a reasonable cost, requiring only a \colp{few matrix multiplications, up to 16} in worst-case scenarios. The method allows us to 
\colp{characterize} the nature of the state and is able to deliver at least a mixed eigenstate when strong degeneracy is encountered. 
Its implementation in DMP-based electronic structure codes or libraries is quite straightforward. An open perspective
is its extension to other interior eigenstates at lower cost using the Lanczos method.

\begin{acknowledgments}
The author gratefully acknowledges T. Prtenjak, Prof. D.  R. Bowler \colp{and the reviewers} 
for their valuable comments and suggestions, which helped improve the manuscript.
\end{acknowledgments}






%


\end{document}